\documentclass[aoas,MSNbibl,nameyear,dvips]{arximspdf}
\usepackage{subenv}
\usepackage{dcolumn}
\usepackage{stfloats}
\usepackage{graphicx}

\doi{10.1214/11-AOAS495}
\volume{5}
\issue{4}
\pubyear{2011}
\firstpage{2403}
\lastpage{2424}

\makeatletter

\fnbelowfloat

\newcolumntype{d}[1]{D{.}{.}{#1}}

\newcommand{\RR}{\mathbf{R}}
\newcommand{\X}{\mathcal{X}}
\newcommand{\HH}{\mathcal{H}}
\newcommand{\R}{\mathcal{R}}
\newcommand{\XX}{\mathbf{R}^p}
\newcommand{\Z}{\mathcal{Z}}
\renewcommand{\P}{\mathcal{P}}
\newcommand{\argmin}{\arg\min}
\newcommand{\argmax}{\arg\max}
\newcommand{\Obj}{\operatorname{Obj}}

\makeatother

\begin{document}
\begin{frontmatter}

\title{Prototype selection for interpretable classification}
\runtitle{Prototype selection}

\begin{aug}
\author[A]{\fnms{Jacob} \snm{Bien}\corref{}\thanksref{t1}\ead[label=e1]{jbien@stanford.edu}}
\and
\author[B]{\fnms{Robert} \snm{Tibshirani}\thanksref{t2}\ead[label=e2]{tibs@stanford.edu}}
\runauthor{J. Bien and R. Tibshirani}
\affiliation{Stanford University}
\dedicated{This paper is dedicated to the memory of Sam Roweis}
\address[A]{Department of Statistics\\
Stanford University\\
Sequoia Hall\\
390 Serra Mall\\
Stanford, California 94305\\
USA\\
\printead{e1}}
\address[B]{Departments\\
\quad of Health, Research, and Policy \\
\quad and Statistics\\
Stanford University\\
Sequoia Hall\\
390 Serra Mall\\
Stanford, California 94305\\
USA\\
\printead{e2}}
\end{aug}

\thankstext{t1}{Supported by the Urbanek Family Stanford
Graduate Fellowship and the Gerald~J.~Lieberman Fellowship.}

\thankstext{t2}{Supported in part by NSF
Grant DMS-99-71405 and National Institutes of Health
Contract N01-HV-28183.}

\received{\smonth{4} \syear{2010}}
\revised{\smonth{5} \syear{2011}}

%
\begin{abstract}
Prototype methods seek a minimal subset of samples that can serve as a
distillation or condensed view of a data set. As the size of modern
data sets grows, being able to present a domain specialist with a short
list of ``representative'' samples chosen from the data set is of
increasing interpretative value. While much recent statistical
research has been focused on producing sparse-in-the-variables methods,
this paper aims at achieving sparsity in the samples.

We discuss a method for selecting prototypes in the classification
setting (in which the samples fall into known discrete categories).
Our method of focus is derived from three basic properties that we
believe a good prototype set should satisfy. This intuition is
translated into a~set cover optimization problem, which we solve
approximately using standard approaches. While prototype selection is
usually viewed as purely a means toward building an efficient
classifier, in this paper we emphasize the inherent value of having a
set of prototypical elements. That said, by using the nearest-neighbor
rule on the set of prototypes, we can of course discuss our method as a
classifier as well.

We demonstrate the interpretative value of producing prototypes on the
well-known USPS ZIP code digits data set and show that as a classifier
it performs reasonably well. We apply the method to a~proteomics data
set in which the samples are strings and therefore not naturally
embedded in a vector space. Our method is compatible with any
dissimilarity measure, making it amenable to situations in which using
a non-Euclidean metric is desirable or even necessary.\looseness=1
\end{abstract}

%

\end{frontmatter}

\section{Introduction}
Much of statistics is based on the notion that averaging over
many elements of a data set is a good thing to do. In this paper, we
take an opposite tack. In certain settings, selecting a
small number of ``representative'' samples from a large data set may be
of greater interpretative value than generating some ``optimal'' linear
combination of all the elements of a
data set. For domain specialists, examining a handful of
representative examples of each class can be highly informative especially
when $n$ is large (since looking through all examples from the original
data set could be overwhelming or even infeasible). Prototype methods
aim to select a relatively small number of
samples from a data set which, if well chosen,
can serve as a summary of the original data set. In this paper, we
motivate a particular method for selecting prototypes in the
classification setting. The resulting method is very similar to Class
Cover Catch Digraphs of \citet{Priebe03}. In fact, we have found many
similar proposals across multiple fields, which we
review later in this paper. What
distinguishes this work from the rest is our interest in prototypes as
a tool for better understanding a data set---that is, making it more
easily ``human-readable.'' The bulk of the previous
literature has been on prototype extraction specifically for building
classifiers. We find it useful to discuss our method as a classifier
to the extent that it permits quantifying its abilities. However, our
primary objective is aiding domain specialists in making sense of their
data sets.

Much recent work in the statistics community has been devoted
to the problem of interpretable classification
through achieving sparsity in the \textit{variables}
[\citet{THNC2002},
\citet{Zhu04}, \citet{PH2006}, \citet{glmnet}]. In
this paper, our aim is
interpretability through sparsity in the \textit{samples}. Consider the
US Postal Service's ZIP code data set, which consists of
a training set of 7,291 grayscale ($16\times16$ pixel) images of
handwritten digits 0--9 with associated labels indicating the intended
digit. A typical ``sparsity-in-the-variables'' method would identify
a subset of the pixels that is most predictive of digit-type. In
contrast, our method identifies a subset of the images that, in a sense,
is most
predictive of digit-type. Figure \ref{fig:digits_protos} shows the
first 88 prototypes selected
by our method. It aims to select prototypes that capture the full
variability of
a~class while avoiding confusion with other classes. For example, it
chooses a~wide enough range of examples of the digit ``7'' to
demonstrate that some people add a~serif while others do
not; however, it avoids any ``7'' examples
that look too much like a ``1.'' We see that many more ``0'' examples
have been chosen
than ``1'' examples despite the fact that the original training set
has roughly the same number of samples of these two classes. This
reflects the fact that there is
much more variability in how people write ``0'' than ``1.''

More generally, suppose we are given a training set of points
$\X=\{\mathbf{x}_1,\ldots,\allowbreak\mathbf{x}_n\}\subset\XX$ with
corresponding class labels
$y_1,\ldots,y_n\in\{1,\ldots,L\}$. The output of our method are
prototype sets $\P_l\subseteq\X$ for each
class $l$. The goal is that someone\vadjust{\goodbreak} given only $\P_1,\ldots,\P_L$
would have a good sense of
the original training data, $\X$~and~$\mathbf{y}$. The above
situation describes the standard setting of a~condensation problem
[\citet{H68}, \citet{Lozano}, \citet{Ripley05}].

At the heart of our proposed method is the premise that the prototypes
of class $l$ should consist of points that are close to many training
points of class $l$ and are far from training points of other classes.
This idea captures the sense in which the word ``prototypical'' is
commonly used.

Besides the interpretative value of prototypes, they also provide a
means for classification. Given the prototype sets $\P_1,\ldots,\P
_L$, we
may classify any new $\mathbf{x}\in\XX$ according to the class whose
$\P_l$ contains
the nearest prototype:
%
%
\begin{equation}\label{eq:chat}
\hat c(\mathbf{x}) = \mathop{\argmin}_{l}\min_{\mathbf{z}\in\P
_l}d(\mathbf
{x},\mathbf{z}).
\end{equation}
Notice that this classification rule reduces to \textit{one nearest
neighbors} (1-NN) in the
case that $\P_l$ consists of all $\mathbf{x}_i\in\X$ with $y_i=l$.

The 1-NN rule's popularity stems from its conceptual simplicity,
empirically good
performance, and theoretical properties [\citet{CH67}]. Nearest
prototype methods seek a lighter-weight representation of the training
set that does not sacrifice (and, in fact, may improve) the accuracy
of the classifier. As a classifier, our method performs reasonably
well, although its main strengths lie in the ease of understanding
why a given prediction has been made---an alternative to (possibly
high-accuracy) ``black box'' methods.

In Section \ref{sec:PVM} we begin with a conceptually simple
optimization criterion that describes a desirable choice for
$\P_1,\ldots,\P_L$. This intuition gives rise to an integer program,
which can be decoupled into $L$ separate set cover problems. In
Section~\ref{sec:solving-problem} we present two approximation
algorithms for solving the optimization problem. Section
\ref{sec:adapt-pvm-spec} discusses considerations for applying our
method most effectively to a given data set. In Section
\ref{sec:related-work} we give an overview of related work. In Section
\ref{sec:exampl-simul-real} we return to the ZIP code digits data set
and present other empirical results, including an application to
proteomics.

\section{Formulation as an optimization problem}
\label{sec:PVM}

In this section we frame prototype selection as an
optimization problem. The problem's connection to set cover will lead
us naturally to an algorithm for prototype selection.

\subsection{The intuition}
\label{sec:intuition}
Our guiding intuition is that a good set of prototypes for class $l$
should capture the full
structure of the training examples of class $l$ while taking into
consideration the structure of other classes. More explicitly, every
training example should have a prototype of its same class in its
neighborhood; no point
should have a prototype of a different class in its neighborhood; and,
finally, there should be as few prototypes as possible.
These three principles capture what we mean by ``prototypical.'' Our method
seeks prototype sets with a slightly relaxed version of these properties.

%
%
\begin{figure}

\includegraphics{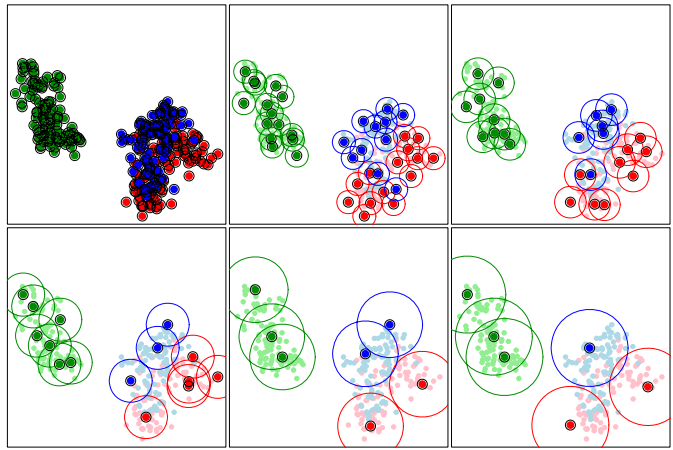}

\caption{Given a value for $\varepsilon$, the choice of
$\P_1,\ldots,\P_L$ induces $L$ partial covers of the training points by
$\varepsilon$-balls. Here $\varepsilon$
is varied from the smallest (top-left panel) to approximately the
median interpoint distance (bottom-right panel).}
\label{fig:demo}
\end{figure}

As a first step, we make the notion of neighborhood more precise. For a~given choice of $\P_l\subseteq\X$, we consider the set of $\varepsilon
$-balls centered at
each $\mathbf{x}_j\in\P_l$ (see Figure \ref{fig:demo}). A desirable
prototype
set for class $l$ is then one that induces a set of balls which:
\begin{longlist}[(a)]
\item[(a)] covers as many training points of
class $l$ as possible,
\item[(b)] covers as few training points as possible of classes other
than $l$, and
\item[(c)] is sparse (i.e., uses as few prototypes as possible for the
given $\varepsilon$).
\end{longlist}

We have thus translated our initial problem concerning prototypes into
the geometric problem
of selectively covering points with a specified set of balls. We will
show that
our problem reduces to the extensively studied set cover problem. We
briefly review set cover before
proceeding with a more precise statement of our problem.

\subsection{The set cover integer program}
Given a set of points $\X$ and a collection of sets that forms a cover
of $\X$, the set cover problem seeks the smallest subcover of~$\X$.
Consider the following special case: Let $B(\mathbf{x}) = \{
\mathbf{x}'\in \RR^p\dvtx d(\mathbf{x}',\mathbf{x}) < \varepsilon\}$
denote the ball of radius $\varepsilon>0$ centered at $\mathbf{x}$
(note: $d$ need not be a metric). Clearly,
$\{B(\mathbf{x}_i)\dvtx\mathbf{x}_i\in\X\}$ is a cover of $\X$. The
goal is to find the smallest subset of points $\P\subseteq\X$ such that
$\{B(\mathbf{x}_i)\dvtx\mathbf {x}_i\in\P\}$ covers~$\X$ (i.e., every
$\mathbf{x}_i\in\X$ is within $\varepsilon$ of some point in $\P$).
This problem can be written as an integer program by introducing
indicator variables: $\alpha_j=1$ if $\mathbf{x}_j \in\P$ and $\alpha_j
= 0$ otherwise. Using this notation, $\sum_{j\dvtx\mathbf{x}_i\in
B(\mathbf{x}_j)}{\alpha_j}$ counts\vspace*{1pt} the number of times $\mathbf {x}_i$
is covered by a $B(\mathbf{x}_j)$ with $\mathbf{x}_j\in\P$. Thus,
requiring that this sum be positive for each $\mathbf{x}_i\in\X $
enforces that $\P$ induces a cover of $\X$. The set cover problem is
therefore equivalent to the following integer program:
%
%
\begin{eqnarray}\label{eq:setcover}
\mbox{minimize }\sum_{j=1}^n \alpha_j \quad\mbox{s.t.}\quad\sum_{j\dvtx\mathbf
{x}_i\in B(\mathbf{x}_j)}{\alpha_j}&\ge&1\qquad \forall\mathbf
{x}_i\in\X,\nonumber\\[-8pt]\\[-8pt]
\alpha_j &\in&\{0,1\}\qquad \forall\mathbf{x}_j\in\X.\nonumber
\end{eqnarray}
A feasible solution to the above integer program is one that has at
least one prototype within $\varepsilon$ of each training point.

Set cover can be seen as a clustering problem in which we wish to find
the smallest number of clusters
such that every point is within $\varepsilon$ of at least one cluster
center. In the language of vector quantization, it seeks the smallest
codebook (restricted to~$\X$) such that no vector is distorted by more
than $\varepsilon$ [\citet{Tipping01}]. It was the use of set
cover in
this context that was the starting point for our work in developing a
prototype method in the classification setting.

\subsection{From intuition to integer program}
\label{sec:statement-problem} We now express the three properties
(a)--(c) in Section \ref {sec:intuition} as an integer program, taking
as a starting point the set cover problem of (\ref{eq:setcover}).
Property (b) suggests that in certain cases it may be necessary to
leave some points of class $l$ uncovered. For this reason, we adopt a
\textit{prize-collecting set cover} framework for our problem, meaning
we assign a cost to each covering set, a penalty for being uncovered to
each point and then find the minimum-cost partial cover
[\citet{Konemann06}]. Let\vspace*{1pt} $\alpha_j^{(l)}\in\{0,1\}$ indicate
whether we choose $\mathbf{x}_j$ to be in $\P_l$ (i.e., to be a
prototype for class $l$). As with set cover, the sum
$\sum_{j\dvtx\mathbf{x}_i\in B(\mathbf{x}_j)}\alpha_j^{(l)}$
counts\vspace*{1pt} the number of balls $B(\mathbf{x}_j)$ with
$\mathbf{x}_j\in\P_l$ that cover the point $\mathbf{x}_i$. We then set
out to solve the following integer program:
\begin{subequation}\label{eq:pvmip}
\begin{eqnarray}
&&\mathop{\mathrm{minimize}}_{\alpha_j^{(l)}, \xi_i, \eta_i}
\sum_{i}{\xi_i}+\sum_{i}{\eta_i}+\lambda\sum_{j,l}{\alpha
_j^{(l)}}\quad\mbox{s.t.}\nonumber\\
\label{first}
&&\quad\sum_{j\dvtx\mathbf{x}_i\in B(\mathbf{x}_j)}{\alpha
_j^{(y_i)}}\ge1-\xi_i\qquad
\forall\mathbf{x}_i\in\X,\\
\label{second}
&&\quad\mathop{\sum_{j\dvtx\mathbf{x}_i\in B(\mathbf{x}_j)}}_{l\ne y_i}\alpha
_j^{(l)}\le
0+\eta_i \qquad\forall\mathbf{x}_i\in\X,\\
&&\quad\alpha_j^{(l)}\in\{0,1\} \qquad\forall j,l,\qquad \xi_i,\eta_i \ge
0 \qquad\forall i\nonumber.
\end{eqnarray}
\end{subequation}
We have introduced two slack variables, $\xi_i$ and
$\eta_i$, per training point $\mathbf{x}_i$. Constraint (\ref{first})
enforces that each training point be covered
by at least one ball of its own class-type\vadjust{\goodbreak} (otherwise $\xi_i=1$).
Constraint (\ref{second})
expresses the condition that training point $\mathbf{x}_i$ not be
covered with balls of
other classes (otherwise $\eta_i>0$). In particular, $\xi_i$ can be
interpreted as indicating whether~$\mathbf{x}_i$ does \textit{not} fall within
$\varepsilon$ of any prototypes of class $y_i$, and $\eta_i$ counts the
number of
prototypes of class other than $y_i$ that are within $\varepsilon$ of
$\mathbf{x}_i$.

Finally, $\lambda\ge0$ is a parameter specifying the cost of adding a
prototype. Its effect is to control the number of prototypes chosen
[corresponding to
property (c) of the last section]. We generally
choose $\lambda=1/n$, so that property (c) serves only as a
``tie-breaker'' for choosing among multiple solutions that
do equally well on properties (a) and (b). Hence, in words, we are
minimizing the sum of (a) the number of points left uncovered, (b) the
number of times a point is wrongly covered, and (c) the number of
covering balls
(multiplied by $\lambda$). The resulting method has a single tuning parameter,
$\varepsilon$ (the ball radius), which can be estimated by cross-validation.

We show in the \hyperref[app]{Appendix} that the above integer program is
equivalent to $L$ separate prize-collecting set
cover problems. Let $\X_l=\{\mathbf{x}_i\in\X\dvtx y_i=l\}$. Then, for each
class $l$, the set $\P_l\subseteq\X$ is given by the solution to
%
%
\begin{eqnarray}\label{eq:pcsc}
&&\mbox{minimize }\sum_{j=1}^m{C_l(j)\alpha_j^{(l)}} + \sum_{\mathbf
{x}_i\in\X_l}{\xi_i}\quad\mbox{s.t.}\nonumber\\
&&\quad\sum_{j\dvtx\mathbf{x}_i\in B(\mathbf{x}_j)}\alpha
_j^{(l)}\ge1-\xi_i
\qquad\forall\mathbf{x}_i\in\X_l,\\
&&\quad\alpha_j^{(l)}\in\{0,1\} \qquad\forall j,\qquad
\xi_i\ge0\qquad\forall
i\dvtx\mathbf{x}_i\in\X_l\nonumber,
\end{eqnarray}
where $C_l(j)= \lambda+ |B(\mathbf{x}_j)\cap(\X\setminus\X_l)|$
is the cost of adding $\mathbf{x}_j$ to $\P_l$ and a~unit
penalty is charged for each point $\mathbf{x}_i$ of class $l$ left uncovered.

\section{Solving the problem: Two approaches}
\label{sec:solving-problem} The prize-collecting set cover problem of
(\ref{eq:pcsc}) can be transformed to a standard set cover problem by
considering each slack variable $\xi_i$ as representing a singleton set
of unit cost [\citet{Konemann06}]. Since set cover is NP-hard, we
do not expect to find a polynomial-time algorithm to solve our problem
exactly. Further, certain inapproximability results have been proven
for the set cover problem
[\citet{Feige98}].\setcounter{footnote}{2}\footnote{We do not
assume in general that the dissimilarities satisfy the triangle
inequality, so we consider arbitrary covering sets.} In what follows,
we present two algorithms for approximately solving our problem, both
based on standard approximation algorithms for set cover.

\subsection{LP relaxation with randomized rounding}
\label{sec:lp-relaxation}
A well-known approach for the set cover problem
is to relax the integer constraints $\alpha_j^{(l)}\in\{0,1\}$ by
replacing it with $0\le
\alpha_j^{(l)}\le1$.\vadjust{\goodbreak} The result is a linear
program (LP), which is convex and easily solved with any LP solver. The
result is subsequently rounded to recover a feasible (though not
necessarily optimal) solution to the original integer program.

Let $\{\alpha_1^{*(l)},\ldots,\alpha_m^{*(l)}\}\cup\{\xi_i^*\dvtx
i\mbox{ s.t. }\mathbf{x}_i\in\X_l\}$ denote a solution to the LP
relaxation of (\ref{eq:pcsc}) with optimal value
$\mathrm{OPT}_{\mathrm{LP}}^{(l)}$.\vspace*{1pt} Since
$\alpha_j^{*(l)},\xi_i^*\in[0,1]$, we may think of these as
probabilities and round each variable to 1 with probability given by
its value in the LP solution. Following \citet{Vazirani01}, we do
this $O({\log}|\X_l|)$ times and take the union of the partial covers
from all iterations.

We apply this randomized rounding technique to approximately solve
(\ref{eq:pcsc}) for each class separately. For class $l$, the
rounding algorithm is as follows:\vspace*{12pt}
\begin{center}
\fbox{
\begin{minipage}{0.95\linewidth}
\small
\begin{itemize}
\item Initialize $A_1^{(l)}=\cdots=A_m^{(l)}=0$ and $S_i=0$
$\forall i\dvtx\mathbf{x}_i\in\X_l$.
\item For $t=1,\ldots,{2\log}|\X_l|$:
\begin{enumerate}[(1)]
\item[(1)] Draw independently $\tilde A_j^{(l)}\sim
\operatorname{Bernoulli}(\alpha_j^{*(l)})$ and $\tilde S_i\sim
\operatorname{Bernoulli}(\xi_i^*)$.
\item[(2)] Update $A_j^{(l)}:= \max(A_j^{(l)},\tilde A_j^{(l)})$ and
$S_i := \max(S_i,\tilde S_i)$.
\end{enumerate}
\item If $\{A_j^{(l)},S_i\}$ is feasible and has objective $\le{2\log}
|\X_l|\mathrm{OPT}_{\mathrm{LP}}^{(l)}$, return $\P_l = \{\mathbf{x}_j\in\X\dvtx
A_j^{(l)}=1\}$.
Otherwise repeat.
\end{itemize}
\end{minipage}
}
\end{center}

\vspace*{12pt}

In practice, we terminate as soon as a feasible solution is achieved. If
after ${2\log}|\X_l|$ steps the solution is still infeasible or the
objective of the rounded solution is more than ${2\log}|\X_l|$ times the
LP objective, then the algorithm is repeated. By the analysis given
in \citet{Vazirani01}, the probability of this happening is less
than $1/2$, so it is unlikely
that we will have to repeat the above algorithm very many times.
Recalling that the LP relaxation gives a lower bound on the integer
program's optimal value, we see that the randomized rounding yields a
$O({\log}|\X_l|)$-factor approximation to (\ref{eq:pcsc}). Doing this
for each class yields overall a $O(K\log N)$-factor approximation to
(\ref{eq:pvmip}), where $N =
{\max_l}|\X_l|$. We can recover the rounded version of the slack
variable $\eta_i$ by $T_i = \sum_{l\neq y_i}\sum_{j\dvtx\mathbf{x}_i\in
B(\mathbf{x}_j)}A_j^{(l)}$.\vspace*{1pt}

One disadvantage of this approach is that it requires
solving an LP, which we have found can be relatively slow and memory-intensive for
large data sets. The approach we describe next is computationally
easier than the LP rounding
method, is deterministic, and provides a natural ordering of the
prototypes. It is thus our preferred method.

\subsection{A greedy approach}
\label{sec:greedy-approach}
Another well-known approximation algorithm for set cover is a greedy approach
[\citet{Vazirani01}]. At each step, the
prototype with the least ratio of cost to number of points
newly covered is added. However, here we present a less standard
greedy algorithm which has certain practical advantages over the
standard one and does not in
our experience do noticeably worse in minimizing the
objective.
At each step we find the $\mathbf{x}_j\in\X$ and class $l$ for which
adding $\mathbf{x}_j$ to $\P_l$ has the best trade-off of covering
previously uncovered training points of
class $l$ while avoiding covering points of other classes.
The incremental improvement of going from $(\P_1,\ldots,\P_L)$ to
$(\P_1,\ldots,\P_{l-1},\P_l\cup\{\mathbf{x}_j\},\P_{l+1},\ldots
,\P_L)$ can be
denoted by $\Delta\Obj(\mathbf{x}_j,l)= \Delta\xi(\mathbf{x}_j,l)
- \Delta\eta(\mathbf{x}_j,l) - \lambda$, where
\begin{eqnarray*}
\Delta\xi(\mathbf{x}_j,l) &=& \biggl|\X_l\cap\biggl(B(\mathbf
{x}_j)\Bigm\backslash
\bigcup_{\mathbf{x}_{j'}\in\P_l}B(\mathbf{x}_{j'})\biggr)\biggr|,\\
\Delta\eta(\mathbf{x}_j,l)&=&|B(\mathbf{x}_j)\cap(\X\setminus\X_l)|.\
\end{eqnarray*}
The greedy algorithm is simply as follows:\vspace*{12pt}
\begin{center}
\fbox{
\begin{minipage}{0.95\linewidth}
\small
\begin{enumerate}[(1)]
\item[(1)] Start with $\P_l=\varnothing$ for each class $l$.
\item[(2)] While $\Delta\Obj(\mathbf{x}^*,l^*)>0$:
\begin{itemize}
\item Find $(\mathbf{x}^*,l^*)=\argmax_{(\mathbf{x}_j,l)}{\Delta
\Obj(\mathbf{x}_j,l)}$.
\item Let $\P_{l^*}:= \P_{l^*}\cup\{\mathbf{x}^*\}$.
\end{itemize}
\end{enumerate}
\end{minipage}
}
\end{center}

\vspace*{12pt}

\begin{figure}

\includegraphics{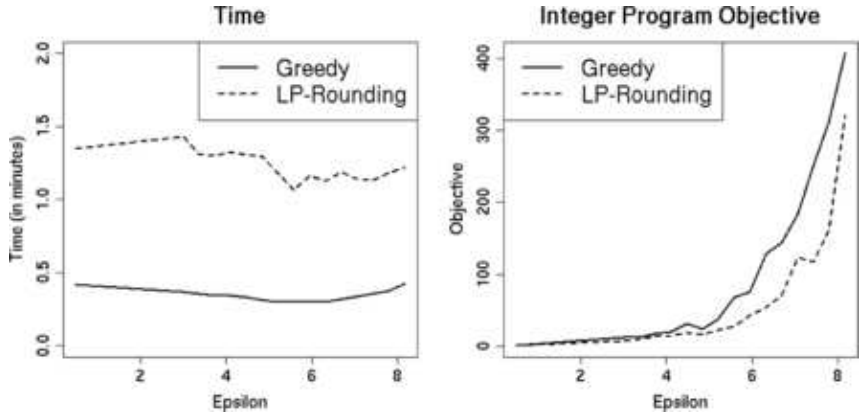}

\caption{Performance comparison of LP-rounding and greedy approaches
on the
digits data set of Section \protect\ref{sec:digits-data}.}
\label{fig:timecompare}
\end{figure}

Figure \ref{fig:timecompare} shows a performance comparison of the two
approaches on the digits data (described in Section
\ref{sec:digits-data}) based on time and resulting (integer program)
objective. Of course, any time comparison is greatly dependent on the
%
%
machine and implementation, and we found great variability in running
time among LP solvers. While low-level, specialized software could lead
to significant time gains, for our present purposes, we use
off-the-shelf, high-level software. The LP was solved using the \texttt
{R} package \texttt{Rglpk}, an interface to the GNU Linear Programming
Kit. For the greedy approach, we wrote a~simple function in \texttt{R}.

\section{Problem-specific considerations}
\label{sec:adapt-pvm-spec}
In this section we describe two ways in which our method can be
tailored by the user for the particular problem at hand.

\subsection{Dissimilarities}
\label{sec:dissimilarities}
Our method depends on the features only through the pairwise
dissimilarities $d(\mathbf{x}_i,\mathbf{x}_j)$, which allows it to
share in the benefits of kernel methods by using a kernel-based
distance. For problems in the $p\gg n$
realm, using distances that effectively lower the dimension can lead
to improvements. Additionally, in problems
in which the data are not readily embedded in a vector space (see
Section \ref{sec:prot-class-with}), our method may still be applied if
pairwise dissimilarities are available. Finally, given any
dissimilarity $d$, we may instead use $\tilde d$,
defined by $\tilde d(\mathbf{x},\mathbf{z}) = |\{\mathbf{x}_i\in\X
\dvtx d(\mathbf{x}_i,\mathbf{z}) \le d(\mathbf{x},\mathbf{z})\}|$.
Using~$\tilde d$ induces $\varepsilon$-balls, $B(\mathbf{x}_j)$, consisting
of the
$(\lfloor\varepsilon\rfloor-1)$ nearest training points to~$\mathbf{x}_j$.

\subsection{Prototypes not on training points}
\label{sec:prot-not-train}
For simplicity, up until now we have described a special case of our
method in
which we only allow prototypes to lie on elements of the training
set $\X$. However, our method is easily generalized to the case where
prototypes are selected from any finite set of points. In particular,
suppose, in addition to the labeled training
data $\X$ and $\mathbf{y}$, we are also given a set $\Z= \{\mathbf
{z}_1,\ldots,\mathbf{z}_m\}$
of unlabeled points. This situation (known as semi-supervised
learning) occurs, for example, when it is expensive to obtain large
amounts of labeled examples, but collecting unlabeled data is cheap.
Taking $\Z$ as the set of potential prototypes, the optimization
problem (\ref{eq:pvmip}) is easily modified so that
$\P_1,\ldots,\P_L$ are selected subsets of $\Z$. Doing so preserves
the property that
all prototypes are actual examples (rather than arbitrary points in
$\XX$).

While having prototypes confined to lie on actual observed
points is desirable for interpretability, if this is not desired, then
$\Z$
may be further augmented to include other points. For example, one
could run
$K$-means on each class's points individually and add these $L\cdot K$
centroids to $\Z$. This method seems to help
especially in high-dimensional problems where constraining all
prototypes to lie on data points suffers from the \textit{curse of
dimensionality}.\looseness=1

\section{Related work}
\label{sec:related-work} Before we proceed with empirical evaluations
of our method, we discuss related work. There is an abundance of
methods that have been proposed addressing the problem of how to select
prototypes from a training set. These proposals appear in multiple
fields under different names and with differing goals and
justifications. The fact that this problem lies at the intersection of
so many different literatures makes it difficult to provide a complete
overview of them all. In some cases, the proposals are quite similar to
our own, differing in minor details or reducing in a special case. What
makes the present work different from the rest is our \textit{goal},
which is to develop an interpretative aid for data analysts who need to
make sense of a large set of labeled data. The details of our method
have been adapted to this goal; however, other proposals---while
perhaps intended specifically as a preprocessing step for the
classification task---may be effectively adapted toward this end as
well. In this section we review some of the related work to our own.

\subsection{Class cover catch digraphs}
\label{sec:class-cover-catch}

Priebe et~al. (\citeyear{Priebe03}) form a directed graph $D_k=(\X_k,
E_k)$ for each
class $k$ where $(\mathbf{x}_i,\mathbf{x}_j)\in E_k$ if a ball
centered at $\mathbf{x}_i$ of radius $r_i$
covers $\mathbf{x}_j$. One choice of $r_i$ is to make it as large as
possible without covering more than a specified number of other-class
points. A dominating set of $D_k$ is a set of nodes for which all
elements of $\X_k$ are reachable by crossing no more than one edge.
They use a greedy algorithm to find an approximation to the minimum
dominating set for each $D_k$. This set of points is then used to
form the Class Cover Catch Digraph (CCCD) Classifier, which is a
nearest neighbor rule that scales distances by the radii. Noting that a
dominating set of $D_k$
corresponds to finding a set of balls that covers all points of class
$k$, we see that their method could also be described in terms of
set cover. The main difference between their formulation and ours is
that we
choose a fixed radius across all points, whereas in their formulation
a large homogeneous region is filled by a large ball. Our choice
of fixed radius seems favorable from an interpretability standpoint since
there can be regions of space which are class-homogeneous and yet for
which there is a~lot of interesting within-class variability which the
prototypes should reveal. The CCCD work is an outgrowth of the
Class Cover Problem, which does not allow balls to cover wrong-class
points [\citet{Cannon04}]. This literature has been developed in
more theoretical directions [e.g., \citet{Devinney02},
\citet{Ceyhan07}].

\subsection{The set covering machine}
Marchand and Shawe-Taylor (\citeyear{SCM}) introduce the \textit{set
covering machine} (SCM) as a method for learning compact disjunctions
(or conjunctions) of $\mathbf{x}$ in the binary classification setting
(i.e., when $L=2$). That is, given a potentially large set of binary
functions of the features, $\HH=\{ h_j,j=1,\ldots,m\}$ where
$h_j\dvtx\XX\to\{0,1\}$, the SCM selects a relatively small subset of
functions, $\R\subseteq\HH$, for which the prediction rule
$f(\mathbf{x}) = \bigvee_{j\in\R} h_j(\mathbf{x})$ (in the case of a
disjunction) has low training error. Although their stated problem is
unrelated to ours, the form of the optimization problem is very
similar.

In \citet{LPSCM} the authors express the SCM optimization problem
explicitly as an integer program, where the binary vector $\alpha$
is of length $m$ and indicates which of the $h_j$ are in $\R$:
%
%
\begin{eqnarray}\label{eq:SCM}
&&\mathop{\mathrm{minimize}}_{\alpha,\xi,\eta} \sum_{j=1}^m \alpha_j + D\Biggl(\sum
_{i=1}^m\xi_i + \sum_{i=1}^m\eta_i\Biggr)\quad\mbox{s.t.}
\nonumber\\[-8pt]\\[-8pt]
&&\quad H_+\alpha\ge1-\xi,\qquad H_-\alpha\le0+\eta,\qquad \alpha
\in\{0,1\}^m;\qquad \xi, \eta\ge0\nonumber.
\end{eqnarray}
In the above integer program (for the disjunction case), $H_+$ is the
matrix with $ij$th entry $h_j(\mathbf{x}_i)$, with each row $i$ corresponding
to a ``positive'' example $\mathbf{x}_i$ and $H_-$ the analogous
matrix for ``negative'' examples. Disregarding the slack vectors $\xi$
and $\eta$,
this seeks the binary vector $\alpha$ for which every positive example
is covered by at least one $h_j\in\R$ and for which no negative
example is covered
by any $h_j\in\R$. The presence of the slack variables permits a certain
number of errors to be made on the training set, with the trade-off
between accuracy and size of $\R$ controlled by the parameter $D$.

A particular choice for $\HH$ is also suggested in \citet{SCM},
which they
call ``data-dependent balls,'' consisting of indicator functions for the
set of all balls with centers at ``positive'' $\mathbf{x}_i$ (and of all
radii) and the complement of all balls centered at ``negative''
$\mathbf{x}_i$.

Clearly, the integer programs (\ref{eq:pvmip}) and (\ref{eq:SCM}) are
very similar. If we
take $\HH$ to be the set of balls of radius $\varepsilon$
with centers at the positive points only, solving~(\ref{eq:SCM}) is
equivalent to finding the set of prototypes for the positive class
using our method. As shown in the \hyperref[app]{Appendix}, (\ref{eq:pvmip})
decouples into $L$ separate problems. Each of these is equivalent to
(\ref{eq:SCM}) with the positive and negative classes being $\X_l$ and
$\X\setminus\X_l$, respectively. Despite this correspondence,
\citet{SCM} were not considering the problem of prototype
selection in
their work. Since Marchand's and Shawe-Taylor's
(\citeyear{SCM}) goal was to learn a
conjunction (or disjunction) of binary features, they take as a~classification rule $f(\mathbf{x})$; since our aim is a set of prototypes,
it is natural that we use the standard nearest-prototype classification
rule of (\ref{eq:chat}).

For solving the SCM integer program, \citet{LPSCM} propose an LP
relaxation; however, a key difference between their approach and ours
is that they do not seek an integer solution (as we do with the
randomized rounding), but rather modify the
prediction rule to make use of the fractional solution directly.

\citet{SCM} propose a greedy approach to solving~(\ref{eq:SCM}).
Our greedy algorithm differs slightly from
theirs in the following respect. In their algorithm, once a point is
misclassified by a feature, no further penalty is incurred for other
features also misclassifying it. In contrast, in our algorithm, a
prototype is always
charged if it falls within $\varepsilon$ of a~wrong-class training
point. This choice is truer to the integer programs (\ref{eq:pvmip})
and (\ref{eq:SCM}) since the objective has $\sum_j \eta_j$ rather than
$\sum_j 1\{\eta_j>0\}$.

\subsection{Condensation and instance selection methods}
Our method (with $\Z=\X$) selects a subset of the original training
set as prototypes. In this
sense, it is similar in spirit to condensing and data editing methods,
such as the \textit{condensed
nearest neighbor rule} [\citet{H68}] and \textit{multiedit}
[\citet{DK82}].
\citet{H68} introduces the notion of the minimal consistent
subset---the smallest subset of $\X$
for which nearest-prototype classification has 0 training error. Our
method's objective,
$\sum_{i=1}^n{\xi_i}+\sum_{i=1}^n{\eta_i}+\lambda\sum
_{j,l}{\alpha_j^{(l)}}$, represents a sort of compromise, governed by
$\lambda$, between consistency
(first two terms) and minimality (third term). In contrast to our
method, which retains examples from the most homogeneous
regions, condensation methods tend to specifically keep those elements
that fall on
the boundary between classes [\citet{Fayed09}]. This difference
highlights the distinction between the goals of reducing a data set for
good classification performance versus creating a tool for
interpreting a data set. \citet{Wilson00} provide a good survey of
\textit{instance-based learning}, focusing---as is typical in this
domain---entirely on
its ability to improve the efficiency and accuracy of classification
rather than discussing its attractiveness for understanding a
data set. More recently, \citet{Cano07} use evolutionary
algorithms to perform instance
selection with the goal of creating decision trees that are both
precise and interpretable, and \citet{Marchiori10} suggests an instance
selection technique focused on having a large hypothesis margin.
\citet{Cano03} compare
the performance of a number of instance selection methods.

\subsection{Other methods}
We also mention a few other nearest prototype methods.
$K$-means and $K$-medoids are common unsupervised methods which produce
prototypes. Simply running these methods on each class separately yields
prototype sets $\P_1,\ldots,\P_L$. $K$-medoids is similar to our
method in that its prototypes are selected from a finite set. In contrast,
$K$-means's prototypes are not required to lie on training points,
making the method \textit{adaptive}. While allowing prototypes to lie
anywhere in $\XX$ can improve classification error, it also reduces
the interpretability of the prototypes (e.g., in data sets where each
$\mathbf{x}_i$
represents an English word, producing a linear combination of
hundreds of words offers little interpretative value). Probably the
most widely used
adaptive prototype method is \textit{learning vector quantization} [LVQ,
\citet{Kohonen01}]. Several versions of LVQ exist, varying in
certain details, but each
begins with an initial set of prototypes and then iteratively adjusts them
in a fashion that tends to encourage each prototype to lie near many
training points of its class and away from training
points of other classes.

\citet{Takigawa} propose an idea similar to ours in which they
select convex sets to represent each class, and then
make predictions for new points by finding the set with
nearest boundary. They refer to the selected convex
sets themselves as prototypes.

Finally, in the main example of this paper (Section \ref
{sec:digits-data}), we
observe that the relative proportion of prototypes selected for each
class reveals that certain classes are far more complex than others.
We note here that quantifying the complexity of a data set is itself a
subject that has
been studied extensively [\citet{Basu06}].

\section{Examples on simulated and real data}
\label{sec:exampl-simul-real}
We demonstrate the use of our method on several data sets and compare
its performance as a classifier to some of the prototype methods
best known to statisticians. Classification error is a convenient metric
for demonstrating that our proposal is reasonable even though building
a classifier is not our focus. All the methods we include are similar
in that they first
choose a set of prototypes and then use the nearest-prototype rule to
classify. LVQ and $K$-means differ from the rest in that they
do not constrain the prototypes to lie on actual elements of the
training set (or any prespecified finite set $\Z$). We view this
flexibility as a~hinderance for interpretability but a potential
advantage for classification error.\looseness=-1

For $K$-medoids, we run the function \texttt{pam} of the \texttt{R}
package \texttt{cluster} on
each class's data separately, producing $K$ prototypes per class. For
LVQ, we use the functions \texttt{lvqinit} and \texttt{olvq1} [optimized
learning vector quantization 1, \citet{Kohonen01}] from the
\texttt{R}
package \texttt{class}. We vary the initial codebook size to produce a
range of
solutions.

\subsection{Mixture of Gaussians simulation}
\label{sec:mixt-gauss-simul}
\begin{figure}[b]

\includegraphics{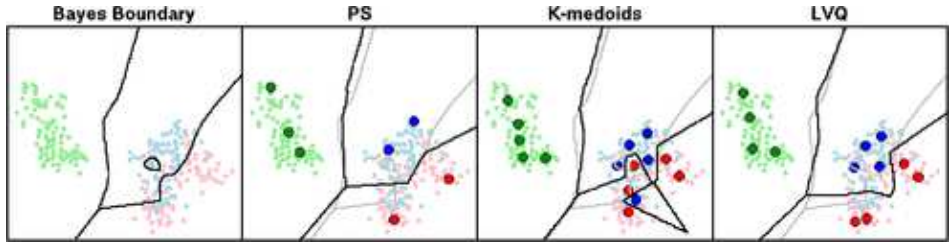}

\caption{Mixture of Gaussians. Classification boundaries of Bayes, our
method (PS), $K$-medoids and LVQ (Bayes boundary in gray for
comparison).} \label{fig:mixture}
\end{figure}

For demonstration purposes, we consider a three-class example with
$p=2$. Each class was generated as a~mixture of 10 Gaussians. Figure
\ref{fig:demo} shows our method's solution
for a range of values of the tuning parameter $\varepsilon$. In Figure
\ref{fig:mixture} we display the classification boundaries of a number
%
%
of methods. Our method (which we label as ``PS,'' for prototype
selection) and LVQ succeed in capturing the shape of the boundary,
whereas $K$-medoids has an erratic boundary; it does not perform well
when classes overlap since it does not take into account other classes
when choosing prototypes.\vadjust{\goodbreak}

\subsection{ZIP code digits data}
\label{sec:digits-data} We return now to the USPS handwritten digits
data set, which consists of a training set of $n=7\mbox{,}291$ grayscale
($16\times16$ pixel) images of handwritten digits 0--9 (and $2\mbox{,}007$ test
images). We ran our method for a range of values of $\varepsilon$ from
the minimum interpoint distance (in which our method retains the entire
training set and so reduces to 1-NN classification) to approximately
the $14$th percentile of interpoint distances.

%
%
\begin{figure}

\includegraphics{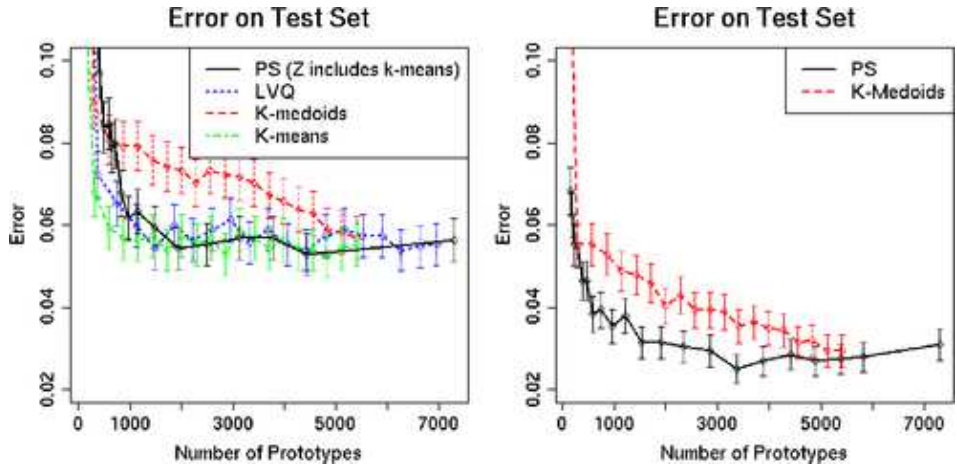}

\caption{Digits data set. Left: all methods use
Euclidean distance and allow prototypes to lie off of training
points (except for $K$-medoids). Right:
both use tangent distance and constrain prototypes to lie on
training points. The rightmost point on our method's
curve (black) corresponds to 1-NN.}
\label{fig:digitstest}
\end{figure}

The left-hand panel of Figure \ref{fig:digitstest} shows the test error
as a function of the number of prototypes for several methods using the
Euclidean metric. Since both LVQ and $K$-means can place prototypes
anywhere in the feature space, which is advantageous in
high-dimensional problems, we also allow our method to select
prototypes that do not lie on the training points by augmenting $\Z$.
In this case, we run 10-means clustering on each class separately and
then add these resulting 100 points to $\Z$ (in addition to $\X$).

The notion of the \textit{tangent distance} between two such images was
introduced by \citet{sim-93} to account for certain invariances
in this problem (e.g., the thickness and orientation of a digit
are not relevant factors when we consider how similar two digits
are). Use of tangent distance with 1-NN attained the lowest test
errors of any method [\citet{HS97}]. Since our method operates on
an arbitrary dissimilarities matrix, we can easily use the tangent distance
in place of the standard Euclidean metric.
The righthand panel of Figure \ref{fig:digitstest} shows the test
errors when tangent distance is used. $K$-medoids similarly
readily accommodates any dissimilarity. While LVQ has been generalized to
arbitrary differentiable metrics, there does not appear to be
generic, off-the-shelf software available. The lowest test error attained
by our method is 2.49\% with a 3,372-prototype solution (compared to 1-NNs
3.09\%).\footnote{\citet{HS97} report a 2.6\% test error for 1-NN
on this data set. The difference may be due to implementation details
of the tangent distance.} Of course, the minimum of the curve is a
biased estimate of test error; however, it is reassuring to note that
for a~wide range of $\varepsilon$ values we get a solution with test
error comparable to that of 1-NN, but requiring far fewer prototypes.

As stated earlier, our primary interest is in the interpretative advantage
offered by our method. A unique feature of our method is that it
automatically chooses
the relative number of prototypes per class to use. In this example, it
is interesting to examine the class-frequencies
of prototypes (Table \ref{tab:numproto}).

%
%
\begin{table}
\caption{Comparison of number of prototypes chosen per
class to training set size}
\label{tab:numproto}
\begin{tabular*}{\tablewidth}{@{\extracolsep{\fill}}lrrccccccccc@{}}
\hline
&\multicolumn{10}{c}{\textbf{Digit}}\\[-4pt]
&\multicolumn{10}{c}{\hrulefill}\\
&\multicolumn{1}{c}{\textbf{0}}&\multicolumn{1}{c}{\textbf{1}}
&\multicolumn{1}{c}{\textbf{2}}&\multicolumn{1}{c}{\textbf{3}}
&\multicolumn{1}{c}{\textbf{4}}&\multicolumn{1}{c}{\textbf{5}}
&\multicolumn{1}{c}{\textbf{6}}&\multicolumn{1}{c}{\textbf{7}}
&\multicolumn{1}{c}{\textbf{8}}&\multicolumn{1}{c}{\textbf{9}}
& \textbf{Total}\\
\hline
Training set&1,194& 1,005& 731& 658& 652& 556& 664& 645& 542&
644& 7,291\\
PS-best& 493& 7& 661& 551& 324& 486& 217& 101& 378& 154& 3,372\\
\hline
\end{tabular*}
\end{table}

The most dramatic feature of this solution is that it only retains seven
of the 1,005 examples of the digit 1. This reflects the fact that,
relative to other digits, the digit 1 has the least variation when
handwritten. Indeed, the average (tangent) distance between digit 1's
in the
training set is less than half that of any other digit (the second
least variable digit is 7). Our choice to force all balls to have the
same radius leads to the property that classes with
greater variability acquire a larger proportion of the prototypes. By
contrast, $K$-medoids requires the user to decide on the relative
proportions of prototypes across the classes.

\begin{figure}

\includegraphics{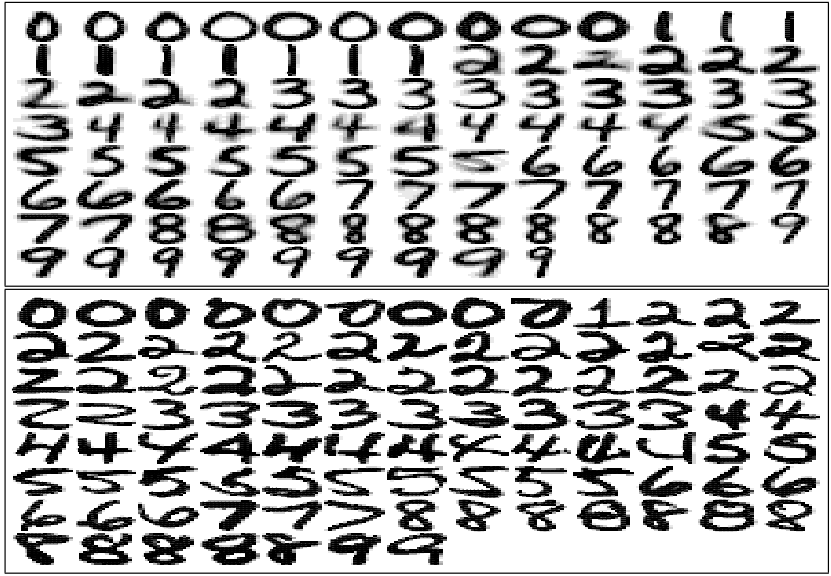}

\caption{(Top) centroids from 10-means clustering within each
class. (Bottom) prototypes from our method (where $\varepsilon$
was chosen to give approximately $100$ prototypes). The images
in the bottom panel are sharper and
show greater variety since each is a single handwritten image.}
\label{fig:blurry}
\end{figure}

Figure \ref{fig:blurry} provides a qualitative comparison between
centroids from $K$-means and prototypes selected by our method. The
upper panel shows the result of 10-means clustering within each class;
the lower panel shows the solution of our method tuned to generate
approximately 100 prototypes. Our prototypes are sharper and show
greater variability than those from $K$-means. Both of these
observations reflect the fact that the $K$-means images are averages of many
training samples, whereas our prototypes are single original images
from the training set. As observed in the 3,372-prototype solution, we
find that the relative numbers of prototoypes in each class for our
method adapts to the within-class variability.

%

%
\begin{figure}[b]

\includegraphics{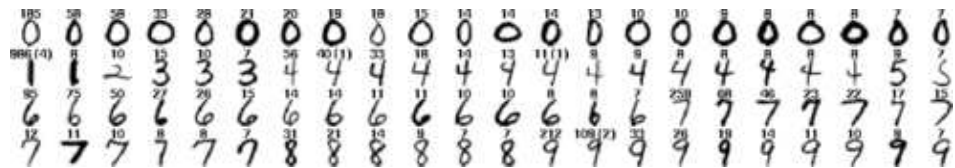}

\caption{First 88 prototypes from greedy algorithm. Above each is
the number of training images first correctly covered by the addition
of this prototype (in parentheses is the
number of miscovered training points by this prototype).}
\label{fig:digits_protos}
\end{figure}

\begin{figure}

\includegraphics{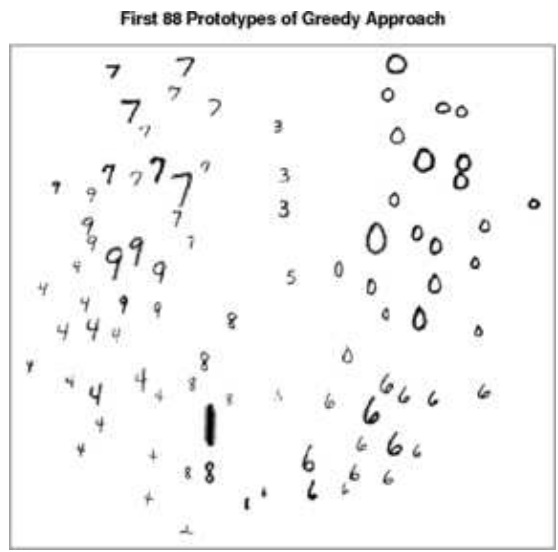}

\caption{The first 88 prototypes (out of 3,372) of the greedy
solution. We perform MDS (\texttt{R} function \texttt{sammon}) on the
tangent distances to visualize the prototypes in two dimensions.
The size of each prototype is proportional to the log of the
number of correct-class training images covered by this prototype.}
\label{fig:88proto_mds}
\end{figure}

\begin{figure}[b]

\includegraphics{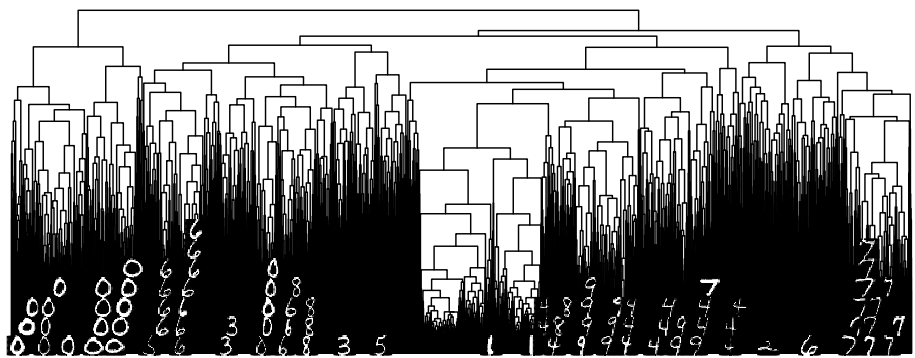}

\caption{Complete-linkage hierarchical clustering of the
training images (using \texttt{R} package \texttt{glus} to order the
leaves). We display the prototype digits where they appear in
the tree. Differing vertical placement of the images is simply to
prevent overlap and has no meaning.}
\label{fig:digits_hclust}
\end{figure}

Figure \ref{fig:digits_protos} shows images of the first 88 prototypes
(of 3,372) selected by the greedy
algorithm. Above each image is the number of training images
previously uncovered that were correctly covered by the addition of
this prototype and, in parentheses, the number of training points that
are miscovered by this prototype. For example, we can
see that the first prototype selected by the greedy algorithm, which
was a
``1,'' covered 986 training images of 1's and four training
images that were not of 1's. Figure \ref{fig:88proto_mds} displays
these in a more visually descriptive way: we have used
multidimensional scaling to arrange the prototypes to reflect the
tangent distances between them. Furthermore, the size of each
prototype is proportional to the log of the number of training images
correctly covered by it. Figure \ref{fig:digits_hclust} shows a~complete-linkage hierarchical clustering of the training set with
images of the 88 prototypes. Figures
\ref{fig:digits_protos}--\ref{fig:digits_hclust} demonstrate ways in
which prototypes can be used to graphically summarize a data set. These
displays could be easily adapted to other domains, for example, by
using gene names in place of the images.

%
%

%
%

\begin{figure}

\includegraphics{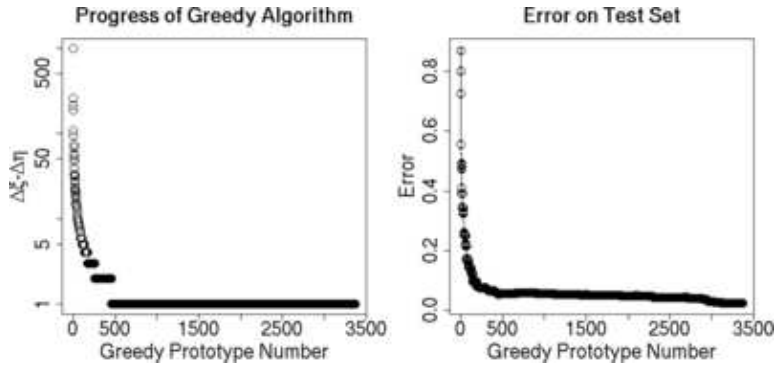}

\caption{Progress of greedy on each iteration.}
\label{fig:greedyscore}
\end{figure}

The left-hand panel of Figure \ref{fig:greedyscore} shows the
improvement in the
objective, $\Delta\xi- \Delta\eta$, after each step of the greedy
algorithm, revealing an interesting feature of the solution: we find
that after the first 458 prototypes are added, each remaining
prototype covers only one training point. Since in this example we
%
%
took $\Z=\X$ (and since a point always covers itself), this means that
the final 2,914 prototypes were chosen to cover only themselves. In
this sense, we see that our method provides a sort of compromise
between a~sparse nearest prototype classifier and 1-NN. This compromise
is determined by the prototype-cost parameter~$\lambda$. If
$\lambda>1$, the algorithm does not enter the 1-NN regime. The
right-hand panel shows that the test error continues to improve as
$\lambda$ decreases.

%
\subsection{Protein classification with string kernels}
\label{sec:prot-class-with}
\begin{figure}

\includegraphics{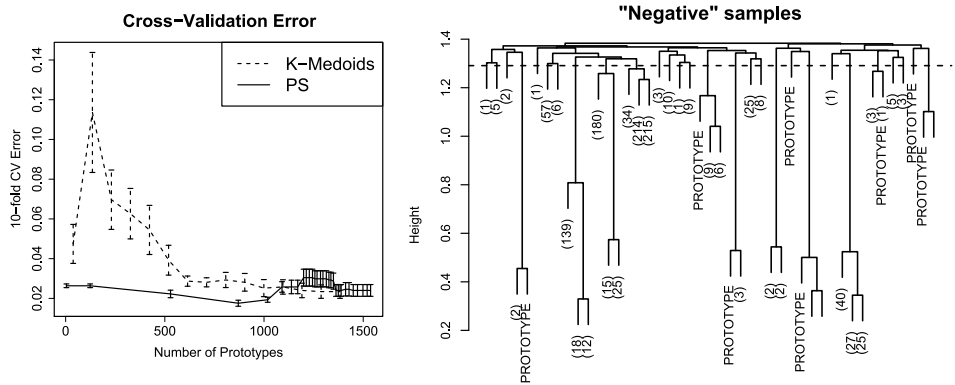}

\caption{Proteins data set. Left: CV error (recall that the
rightmost point on
our method's
curve corresponds to 1-NN). Right: a complete-linkage
hierarchical clustering of the negative samples. Each selected prototype
is marked. The dashed line is a cut at height
$\varepsilon$. Thus, samples that are merged below this line are
within $\varepsilon$ of each other. The number of ``positive'' samples
within $\varepsilon$ of each negative sample, if nonzero, is shown in
parentheses.}
\label{fig:stringkerneltest}
\end{figure}

We next present a case in which the training samples are not
naturally represented as vectors in $\RR^p$. \citet{Leslie04}
study the problem of classification of proteins based on their
amino acid sequences. They introduce a measure of similarity between
protein sequences called the \textit{mismatch kernel}. The general idea
is that two
sequences should be considered similar if they have a large number of
short sequences in common (where two short sequences are considered
the same if they have no more than a specified number of mismatches).
We take as input a $1\mbox{,}708\times1\mbox{,}708$ matrix with $K_{ij}$ containing
the value of the normalized mismatch kernel evaluated between proteins
$i$ and $j$ [the data and software are from \citet{Leslie04}].
The proteins fall into two classes, ``Positive'' and
``Negative,'' according to whether they belong to a certain protein
family. We compute pairwise distances from this kernel via $D_{ij} =
\sqrt{K_{ii}+K_{jj}-2K_{ij}}$ and then\vspace*{1pt} run our method and $K$-medoids.
The left panel of Figure \ref{fig:stringkerneltest}
shows the 10-fold cross-validated errors for our method and
%
%
$K$-medoids. For
our method, we take a range of equally-spaced quantiles of the pairwise
distances from the minimum to the median for the parameter
$\varepsilon$. For $K$-medoids, we take as parameter the fraction of
proteins in each class that should be prototypes. This choice of
parameter allows the classes to have different numbers of prototypes,
which is important in this example because the classes are
greatly imbalanced (only 45 of the 1,708 proteins are in class
``Positive''). The right panel of Figure \ref{fig:stringkerneltest}
shows a complete linkage hierarchical clustering of the 45 samples in
the ``Negative'' class with the selected prototypes indicated. Samples
joined below the dotted line are
within $\varepsilon$ of each other. Thus, performing regular set cover
would result in every branch that is cut at this height having at least
one prototype sample
selected. By contrast, our method leaves some branches without
prototypes. In parentheses, we display the number of samples from the
``Positive'' class that are within~$\varepsilon$ of each ``Negative''
sample. We see that the branches that do not have protoypes are those
for which every ``Negative'' sample has too many ``Positive'' samples within
$\varepsilon$ to make it a worthwhile addition to the prototype
set.\looseness=1

The minimum CV-error (1.76\%) is attained by our method using
about 870 prototypes (averaged over the 10 models fit for that value
of $\varepsilon$). This error is identical to the minimum CV-error of a
support vector machine (tuning the cost parameter) trained using this kernel.
Fitting a model to the whole data set with the selected value of
$\varepsilon$, our method chooses 26 prototypes (of 45) for class
``Positive'' and 907 (of
1,663) for class ``Negative.''

\subsection{UCI data sets}
\label{sec:uci-data-sets}
Finally, we run our method on six data sets from the UCI Machine Learning
Repository [\citet{UCI}] and compare its performance to that of 1-NN
(i.e., retaining all training points as prototypes), \mbox{$K$-medoids}
and LVQ. We randomly select $2/3$ of each data set for training and use
the remainder as a test set. Ten-fold cross-validation [and the
``1 standard error
rule,'' \citet{ESL}] is performed on the training data to select a
value for each method's tuning parameter (except for 1-NN). Table
\ref{tab:uci} reports the error on the test set and the number of
prototypes selected for each method. For methods taking a~%
%
\begin{table}
\tabcolsep=0pt
\caption{10-fold CV (with the 1 SE rule) on the training
set to tune
the parameters (our method labeled ``PS'')}
\label{tab:uci}
\begin{tabular*}{\tablewidth}{@{\extracolsep{\fill}}lcd{3.1}d{3.1}d{3.1}d{3.1}d{3.1}d{3.1}
d{3.1}@{}}
\hline
\textbf{Data}&&\multicolumn{1}{c}{$\mbox{\textbf{1-NN}}\bolds{/\ell_2}$}
& \multicolumn{1}{c}{$\mbox{\textbf{1-NN}}\bolds{/\ell
_1}$} & \multicolumn{1}{c}{$\mbox{\textbf{PS}}\bolds{/\ell_2}$}
& \multicolumn{1}{c}{$\mbox{\textbf{PS}}\bolds{/\ell_1}$}
& \multicolumn{1}{c}{$\bolds{K}\mbox{\textbf{-med.}}\bolds{/\ell_2}$}
& \multicolumn{1}{c}{$\bolds{K}\mbox{\textbf{-med.}}\bolds{/\ell_1}$}
& \multicolumn{1}{c@{}}{\textbf{LVQ}}\\
\hline
Diabetes&\textit{Test Err}&28.9&31.6&24.2&26.6&32.0&34.4&25.0\\
($p=8,L=2$)&\textit{\# Protos}&512&512&12&5&194&60&29\\
[4pt]
Glass&\textit{Test Err}&38.0&32.4&36.6&47.9&39.4&38.0&35.2\\
($p=9,L=6$)&\textit{\# Protos}&143&143&34&17&12&24&17\\
[4pt]
Heart&\textit{Test Err}&21.1&23.3&21.1&13.3&22.2&24.4&15.6\\
($p=13,L=2$)&\textit{\# Protos}&180&180&6&4&20&20&12\\
[4pt]
Liver&\textit{Test Err}&41.7&41.7&41.7&32.2&46.1&48.7&33.9\\
($p=6,L=2$)&\textit{\# Protos}&230&230&16&13&120&52&110\\
[4pt]
Vowel&\textit{Test Err}&2.8&1.7&2.8&1.7&2.8&4.0&24.4\\
($p=10,L=11$)&\textit{\# Protos}&352&352&352&352&198&165&138\\
[4pt]
Wine&\textit{Test Err}&3.4&3.4&11.9&6.8&6.8&1.7&3.4\\
($p=13,L=3$)&\textit{\# Protos}&119&119&4&3&12&39&3\\
\hline
\end{tabular*}
\end{table}
dissimilarity matrix as input, we use both $\ell_2$ and $\ell_1$
distance measures. We see that in most cases our method is able to do as
well as or better than 1-NN but with a significant reduction in
prototypes. No single method does best on all of the data sets. The
difference in results observed for using $\ell_1$ versus $\ell_2$
distances reminds us that the choice of dissimilarity is an important
aspect of any problem.

%
\section{Discussion}
We have presented a straightforward procedure for selecting
prototypical samples from a data set, thus providing a simple way to
``summarize'' a data set. We began by explicitly laying out our notion
of a~desirable
prototype set, then cast this intuition as a set cover problem which
led us to
two standard approximation algorithms. The digits data example
highlights several strengths. Our method automatically chooses a~suitable number of prototypes for each class. It is flexible in that
it can be used in conjunction with a problem-specific dissimilarity,
which in this case helps our method attain a competitive test error
for a
wide range of values of the tuning parameter. However, the main
motivation for using this method is interpretability: each prototype
is an element of $\X$ (i.e., is an actual hand drawn image). In medical
applications, this would mean that prototypes correspond to actual
patients, genes, etc. This feature should be useful to domain
experts for making sense of large data sets. Software for our method
will be made available as an \texttt{R} package in the \texttt{R} library.

\begin{appendix}\label{app}
\section*{\texorpdfstring{Appendix: Integer program (\lowercase{\protect\ref{eq:pvmip}})'s relation to
    prize-collecting set cover}
{Appendix: Integer program (3)'s relation to
    prize-collecting set cover}}

\begin{Claim*} Solving the integer program of
(\ref{eq:pvmip}) is equivalent to solving $L$ prize-collecting set
cover problems.
\end{Claim*}
\begin{pf}
Observing that the constraints (\ref{second}) are always tight, we can
eliminate $\eta_1,\ldots,\eta_n$ in (\ref{eq:pvmip}), yielding
\begin{eqnarray*}
&&\mathop{\mathrm{minimize}}_{\alpha_j^{(l)}, \xi_i, \eta_i}
\sum_{i}{\xi_i}+\sum_{i}{\mathop{\sum_{j\dvtx\mathbf{x}_i\in
B(\mathbf{z}_j)}}_{l\ne y_i}\alpha_j^{(l)}}+\lambda\sum
_{j,l}{\alpha_j^{(l)}}\quad\mbox{s.t.}\\
&&\quad \sum_{j\dvtx\mathbf{x}_i\in B(\mathbf{z}_j)}{\alpha
_j^{(y_i)}}\ge1-\xi_i\qquad
\forall\mathbf{x}_i\in\X,\\
&&\quad\alpha_j^{(l)}\in\{0,1\} \qquad\forall j,l,\qquad \xi_i\ge0 \qquad\forall
i.
\end{eqnarray*}
Rewriting the second term of the objective as
\begin{eqnarray*}
\sum_{i=1}^n{\mathop{\sum_{j\dvtx\mathbf{x}_i\in B(\mathbf
{z}_j)}}_{l\ne
y_i}\alpha_j^{(l)}}
&=& \sum_{j,l}\alpha_j^{(l)}\sum_{i=1}^n 1\{\mathbf{x}_i \in
B(\mathbf{z}_j),\mathbf{x}_i\notin\X_l\}\\
&=& \sum_{j,l}\alpha_j^{(l)}|B(\mathbf{z}_j)\cap(\X\setminus\X_l)|
\end{eqnarray*}
and letting $C_l(j) = \lambda+ |B(\mathbf{z}_j)\cap(\X\setminus\X
_l)|$ gives
\[
\mathop{\mathrm{minimize}}_{\alpha_j^{(l)}, \xi_i}
\sum_{l=1}^L{\Biggl[\sum_{\mathbf{x}_i\in\X_l}{\xi_i}+\sum
_{j=1}^m{ C_l(j)\alpha_j^{(l)}}\Biggr]}
\]
s.t. for each class $l$:
\begin{eqnarray*}
&&\sum_{j\dvtx\mathbf{x}_i\in B(\mathbf{z}_j)}{\alpha_j^{(l)}}\ge1-\xi_i
\qquad\forall\mathbf{x}_i\in\X_l,\\
&&\alpha_j^{(l)}\in\{0,1\} \qquad\forall j,\qquad \xi_i\ge0 \qquad\forall
i\dvtx\mathbf{x}_i\in\X_l.
\end{eqnarray*}
This is separable with respect
to class and thus equivalent to $L$ separate integer programs. The
$l$th integer program has variables $\alpha_1^{(l)},\ldots
,\alpha_m^{(l)}$ and
$\{\xi_i\dvtx\mathbf{x}_i\in\X_l\}$ and is precisely the
prize-collecting set
cover problem of (\ref{eq:pcsc}).
\end{pf}
\end{appendix}

\section*{Acknowledgments}

We thank Sam Roweis for showing us set cover as a~clustering method,
Sam Roweis, Amin Saberi, Daniela Witten for helpful discussions, and
Trevor Hastie for providing us with his code for computing tangent
distance.


%

\printaddresses


\begin{thebibliography}{33}

\bibitem[\protect\citeauthoryear{Asuncion and Newman}{2007}]{UCI}
%
\begin{bmisc}[author]
\bauthor{\bsnm{Asuncion},~\bfnm{A.}\binits{A.}} \AND
\bauthor{\bsnm{Newman},~\bfnm{D.~J.}\binits{D.~J.}}
(\byear{2007}).
\bhowpublished{UCI Machine Learning Repository.
Univ. California, Irvine, School of Information and Computer
Sciences.}
\bptok{imsref}%
\end{bmisc}
%
\endbibitem

\bibitem[\protect\citeauthoryear{Basu and Ho}{2006}]{Basu06}
%
\begin{bbook}[author]
\bauthor{\bsnm{Basu},~\bfnm{M.}\binits{M.}} \AND
\bauthor{\bsnm{Ho},~\bfnm{T.~K.}\binits{T.~K.}}
(\byear{2006}).
\btitle{Data Complexity in Pattern Recognition}.
\bpublisher{Springer}, \baddress{London}.
\bptok{imsref}%
\end{bbook}
%
\endbibitem

\bibitem[\protect\citeauthoryear{Cannon and Cowen}{2004}]{Cannon04}
%
\begin{barticle}[mr]
\bauthor{\bsnm{Cannon},~\bfnm{Adam~H.}\binits{A.~H.}} \AND
\bauthor{\bsnm{Cowen},~\bfnm{Lenore~J.}\binits{L.~J.}}
(\byear{2004}).
\btitle{Approximation algorithms for the class cover problem}.
\bjournal{Ann. Math. Artif. Intell.}
\bvolume{40}
\bpages{215--223}.
\bid{doi={10.1023/B:AMAI.0000012867.03976.a5}, issn={1012-2443}, mr={2037478}}
\bptok{imsref}%
\end{barticle}
%
\endbibitem

\bibitem[\protect\citeauthoryear{Cano, Herrera and Lozano}{2003}]{Cano03}
%
\begin{barticle}[author]
\bauthor{\bsnm{Cano},~\bfnm{J.~R.}\binits{J.~R.}},
\bauthor{\bsnm{Herrera},~\bfnm{F.}\binits{F.}} \AND
\bauthor{\bsnm{Lozano},~\bfnm{M.}\binits{M.}}
(\byear{2003}).
\btitle{{Using evolutionary algorithms as instance selection for data reduction
in KDD: An experimental study}}.
\bjournal{IEEE Transactions on Evolutionary Computation}
\bvolume{7}
\bpages{561--575}.
\bptok{imsref}%
\end{barticle}
%
\endbibitem

\bibitem[\protect\citeauthoryear{Cano, Herrera and Lozano}{2007}]{Cano07}
%
\begin{barticle}[author]
\bauthor{\bsnm{Cano},~\bfnm{J.~R.}\binits{J.~R.}},
\bauthor{\bsnm{Herrera},~\bfnm{F.}\binits{F.}} \AND
\bauthor{\bsnm{Lozano},~\bfnm{M.}\binits{M.}}
(\byear{2007}).
\btitle{{Evolutionary stratified training set selection for extracting
classification rules with trade off precision-interpretability}}.
\bjournal{Data and Knowledge Engineering}
\bvolume{60}
\bpages{90--108}.
\bptok{imsref}%
\end{barticle}
%
\endbibitem

\bibitem[\protect\citeauthoryear{Ceyhan, Priebe and Marchette}{2007}]{Ceyhan07}
%
\begin{barticle}[mr]
\bauthor{\bsnm{Ceyhan},~\bfnm{Elvan}\binits{E.}},
\bauthor{\bsnm{Priebe},~\bfnm{Carey~E.}\binits{C.~E.}} \AND
\bauthor{\bsnm{Marchette},~\bfnm{David~J.}\binits{D.~J.}}
(\byear{2007}).
\btitle{A new family of random graphs for testing spatial segregation}.
\bjournal{Canad. J. Statist.}
\bvolume{35}
\bpages{27--50}.
\bid{doi={10.1002/cjs.5550350106}, issn={0319-5724}, mr={2345373}}
\bptok{imsref}%
\end{barticle}
%
\endbibitem

\bibitem[\protect\citeauthoryear{Cover and Hart}{1967}]{CH67}
%
\begin{barticle}[author]
\bauthor{\bsnm{Cover},~\bfnm{T.~M.}\binits{T.~M.}} \AND
\bauthor{\bsnm{Hart},~\bfnm{P.}\binits{P.}}
(\byear{1967}).
\btitle{Nearest neighbor pattern classification}.
\bjournal{Proc. IEEE Trans. Inform. Theory}
\bvolume{IT-11}
\bpages{21--27}.
\bptok{imsref}%
\end{barticle}
%
\endbibitem

\bibitem[\protect\citeauthoryear{Devijver and Kittler}{1982}]{DK82}
%
\begin{bbook}[mr]
\bauthor{\bsnm{Devijver},~\bfnm{Pierre~A.}\binits{P.~A.}} \AND
\bauthor{\bsnm{Kittler},~\bfnm{Josef}\binits{J.}}
(\byear{1982}).
\btitle{Pattern Recognition: A Statistical Approach}.
\bpublisher{Prentice Hall}, \baddress{Englewood Cliffs,
NJ}.
\bid{mr={0692767}}
\bptok{imsref}%
\end{bbook}
%
\endbibitem

\bibitem[\protect\citeauthoryear{DeVinney and Wierman}{2002}]{Devinney02}
%
\begin{barticle}[mr]
\bauthor{\bsnm{DeVinney},~\bfnm{Jason}\binits{J.}} \AND
\bauthor{\bsnm{Wierman},~\bfnm{John~C.}\binits{J.~C.}}
(\byear{2002}).
\btitle{A {SLLN} for a one-dimensional class cover problem}.
\bjournal{Statist. Probab. Lett.}
\bvolume{59}
\bpages{425--435}.
\bid{doi={10.1016/S0167-7152(02)00243-2}, issn={0167-7152}, mr={1935677}}
\bptok{imsref}%
\end{barticle}
%
\endbibitem

\bibitem[\protect\citeauthoryear{Fayed and Atiya}{2009}]{Fayed09}
%
\begin{barticle}[author]
\bauthor{\bsnm{Fayed},~\bfnm{H.~A.}\binits{H.~A.}} \AND
\bauthor{\bsnm{Atiya},~\bfnm{A.~F.}\binits{A.~F.}}
(\byear{2009}).
\btitle{A novel template reduction approach for the $K$-nearest neighbor
method}.
\bjournal{IEEE Transactions on Neural Networks}
\bvolume{20}
\bpages{890--896}.
\bptok{imsref}%
\end{barticle}
%
\endbibitem

\bibitem[\protect\citeauthoryear{Feige}{1998}]{Feige98}
%
\begin{barticle}[mr]
\bauthor{\bsnm{Feige},~\bfnm{Uriel}\binits{U.}}
(\byear{1998}).
\btitle{A threshold of {$\ln n$} for approximating set cover}.
\bjournal{J. ACM}
\bvolume{45}
\bpages{634--652}.
\bid{doi={10.1145/285055.285059}, issn={0004-5411}, mr={1675095}}
\bptok{imsref}%
\end{barticle}
%
\endbibitem

\bibitem[\protect\citeauthoryear{Friedman, Hastie and
Tibshirani}{2010}]{glmnet}
%
\begin{barticle}[author]
\bauthor{\bsnm{Friedman},~\bfnm{Jerome~H.}\binits{J.~H.}},
\bauthor{\bsnm{Hastie},~\bfnm{Trevor}\binits{T.}} \AND
\bauthor{\bsnm{Tibshirani},~\bfnm{Rob}\binits{R.}}
(\byear{2010}).
\btitle{Regularization paths for generalized linear models via coordinate
descent}.
\bjournal{Journal of Statistical Software}
\bvolume{33}
\bpages{1--22}.
\bptok{imsref}%
\end{barticle}
%
\endbibitem

\bibitem[\protect\citeauthoryear{Hart}{1968}]{H68}
%
\begin{barticle}[author]
\bauthor{\bsnm{Hart},~\bfnm{P.}\binits{P.}}
(\byear{1968}).
\btitle{The condensed nearest-neighbor rule}.
\bjournal{IEEE Trans. Inform. Theory}
\bvolume{14}
\bpages{515--516}.
\bptok{imsref}%
\end{barticle}
%
\endbibitem

\bibitem[\protect\citeauthoryear{Hastie and Simard}{1998}]{HS97}
%
\begin{barticle}[author]
\bauthor{\bsnm{Hastie},~\bfnm{T.}\binits{T.}} \AND
\bauthor{\bsnm{Simard},~\bfnm{P.~Y.}\binits{P.~Y.}}
(\byear{1998}).
\btitle{Models and metrics for handwritten digit recognition}.
\bjournal{Statist. Sci.}
\bvolume{13}
\bpages{54--65}.
\bptok{imsref}%
\end{barticle}
%
\endbibitem

\bibitem[\protect\citeauthoryear{Hastie, Tibshirani and Friedman}{2009}]{ESL}
%
\begin{bbook}[mr]
\bauthor{\bsnm{Hastie},~\bfnm{Trevor}\binits{T.}},
\bauthor{\bsnm{Tibshirani},~\bfnm{Robert}\binits{R.}} \AND
\bauthor{\bsnm{Friedman},~\bfnm{Jerome}\binits{J.}}
(\byear{2009}).
\btitle{The Elements of Statistical Learning: Data Mining, Inference, and Prediction},
\bedition{2nd} ed.
\bpublisher{Springer}, \baddress{New York}.
\bid{doi={10.1007/978-0-387-84858-7}, mr={2722294}}
\bptok{imsref}%
\end{bbook}
%
\endbibitem

\bibitem[\protect\citeauthoryear{Hussain, Szedmak and
Shawe-Taylor}{2004}]{LPSCM}
%
\begin{bmisc}[author]
\bauthor{\bsnm{Hussain},~\bfnm{Z.}\binits{Z.}},
\bauthor{\bsnm{Szedmak},~\bfnm{S.}\binits{S.}} \AND
\bauthor{\bsnm{Shawe-Taylor},~\bfnm{J.}\binits{J.}}
(\byear{2004}).
\bhowpublished{The linear programming set covering machine.
\textit{Pattern Analysis,
Statistical Modelling and Computational Learning}.}
\bptok{imsref}%
\end{bmisc}
%
\endbibitem

\bibitem[\protect\citeauthoryear{Kohonen}{2001}]{Kohonen01}
%
\begin{bbook}[mr]
\bauthor{\bsnm{Kohonen},~\bfnm{Teuvo}\binits{T.}}
(\byear{2001}).
\btitle{Self-Organizing Maps},
\bedition{3rd} ed.
\bseries{Springer Series in Information Sciences}
\bvolume{30}.
\bpublisher{Springer}, \baddress{Berlin}.
\bid{mr={1844512}}
\bptok{imsref}%
\end{bbook}
%
\endbibitem

\bibitem[\protect\citeauthoryear{K{\"o}nemann, Parekh and
Segev}{2006}]{Konemann06}
%
\begin{bincollection}[mr]
\bauthor{\bsnm{K{\"o}nemann},~\bfnm{Jochen}\binits{J.}},
\bauthor{\bsnm{Parekh},~\bfnm{Ojas}\binits{O.}} \AND
\bauthor{\bsnm{Segev},~\bfnm{Danny}\binits{D.}}
(\byear{2006}).
\btitle{A unified approach to approximating partial covering problems}.
In \bbooktitle{Algorithms---{ESA} 2006}.
\bseries{Lecture Notes in Computer Science}
\bvolume{4168}
\bpages{468--479}.
\bpublisher{Springer}, \baddress{Berlin}.
\bid{doi={10.1007/11841036_43}, mr={2347166}}
\bptok{imsref}%
\end{bincollection}
%
\endbibitem

\bibitem[\protect\citeauthoryear{Leslie et~al.}{2004}]{Leslie04}
%
\begin{barticle}[pbm]
\bauthor{\bsnm{Leslie},~\bfnm{Christina~S.}\binits{C.~S.}},
\bauthor{\bsnm{Eskin},~\bfnm{Eleazar}\binits{E.}},
\bauthor{\bsnm{Cohen},~\bfnm{Adiel}\binits{A.}},
\bauthor{\bsnm{Weston},~\bfnm{Jason}\binits{J.}} \AND
\bauthor{\bsnm{Noble},~\bfnm{William~Stafford}\binits{W.~S.}}
(\byear{2004}).
\btitle{Mismatch string kernels for discriminative protein classification}.
\bjournal{Bioinformatics}
\bvolume{20}
\bpages{467--476}.
\bid{doi={10.1093/bioinformatics/btg431}, issn={1367-4803}, pii={btg431},
pmid={14990442}}
\bptok{imsref}%
\end{barticle}
%
\endbibitem

\bibitem[\protect\citeauthoryear{Lozano et~al.}{2006}]{Lozano}
%
\begin{barticle}[author]
\bauthor{\bsnm{Lozano},~\bfnm{M.}\binits{M.}},
\bauthor{\bsnm{Sotoca},~\bfnm{J.~M.}\binits{J.~M.}},
\bauthor{\bsnm{S{\'{a}}nchez},~\bfnm{J.~S.}\binits{J.~S.}},
\bauthor{\bsnm{Pla},~\bfnm{F.}\binits{F.}},
\bauthor{\bsnm{Pkalska},~\bfnm{E.}\binits{E.}} \AND
\bauthor{\bsnm{Duin},~\bfnm{R.~P.~W.}\binits{R.~P.~W.}}
(\byear{2006}).
\btitle{Experimental study on prototype optimisation algorithms for
prototype-based classification in vector spaces}.
\bjournal{Pattern Recognition}
\bvolume{39}
\bpages{1827--1838}.
\bptok{imsref}%
\end{barticle}
%
\endbibitem

\bibitem[\protect\citeauthoryear{Marchand and Shawe-Taylor}{2002}]{SCM}
%
\begin{barticle}[author]
\bauthor{\bsnm{Marchand},~\bfnm{M.}\binits{M.}} \AND
\bauthor{\bsnm{Shawe-Taylor},~\bfnm{J.}\binits{J.}}
(\byear{2002}).
\btitle{The set covering machine}.
\bjournal{J. Mach. Learn. Res.}
\bvolume{3}
\bpages{723--746}.
\bptok{imsref}%
\end{barticle}
%
\endbibitem

\bibitem[\protect\citeauthoryear{Marchiori}{2010}]{Marchiori10}
%
\begin{barticle}[pbm]
\bauthor{\bsnm{Marchiori},~\bfnm{Elena}\binits{E.}}
(\byear{2010}).
\btitle{Class conditional nearest neighbor for large margin instance
selection}.
\bjournal{IEEE Trans. Pattern Anal. Mach. Intell.}
\bvolume{32}
\bpages{364--370}.
\bid{doi={10.1109/TPAMI.2009.164}, issn={1939-3539}, pmid={20075464}}
\bptok{imsref}%
\end{barticle}
%
\endbibitem

\bibitem[\protect\citeauthoryear{Park and Hastie}{2007}]{PH2006}
%
\begin{barticle}[mr]
\bauthor{\bsnm{Park},~\bfnm{Mee~Young}\binits{M.~Y.}} \AND
\bauthor{\bsnm{Hastie},~\bfnm{Trevor}\binits{T.}}
(\byear{2007}).
\btitle{{$L\sb1$}-regularization path algorithm for generalized linear
models}.
\bjournal{J. R. Stat. Soc. Ser. B Stat. Methodol.}
\bvolume{69}
\bpages{659--677}.
\bid{doi={10.1111/j.1467-9868.2007.00607.x}, issn={1369-7412}, mr={2370074}}
\bptok{imsref}%
\end{barticle}
%
\endbibitem

\bibitem[\protect\citeauthoryear{Priebe et~al.}{2003}]{Priebe03}
%
\begin{barticle}[mr]
\bauthor{\bsnm{Priebe},~\bfnm{Carey~E.}\binits{C.~E.}},
\bauthor{\bsnm{DeVinney},~\bfnm{Jason~G.}\binits{J.~G.}},
\bauthor{\bsnm{Marchette},~\bfnm{David~J.}\binits{D.~J.}} \AND
\bauthor{\bsnm{Socolinsky},~\bfnm{Diego~A.}\binits{D.~A.}}
(\byear{2003}).
\btitle{Classification using class cover catch digraphs}.
\bjournal{J. Classification}
\bvolume{20}
\bpages{3--23}.
\bid{doi={10.1007/s00357-003-0003-7}, issn={0176-4268}, mr={1983119}}
\bptok{imsref}%
\end{barticle}
%
\endbibitem

\bibitem[\protect\citeauthoryear{Ripley}{2005}]{Ripley05}
%
\begin{bbook}[author]
\bauthor{\bsnm{Ripley},~\bfnm{B.~D.}\binits{B.~D.}}
(\byear{2005}).
\btitle{Pattern Recognition and Neural Networks}.
\bpublisher{Cambridge Univ. Press}, \baddress{New York}.
\bptok{imsref}%
\end{bbook}
%
\endbibitem

\bibitem[\protect\citeauthoryear{Simard, Le~Cun and Denker}{1993}]{sim-93}
%
\begin{binproceedings}[author]
\bauthor{\bsnm{Simard},~\bfnm{P.~Y.}\binits{P.~Y.}},
\bauthor{\bsnm{Le~Cun},~\bfnm{Y.~A.}\binits{Y.~A.}} \AND
\bauthor{\bsnm{Denker},~\bfnm{J.~S.}\binits{J.~S.}}
(\byear{1993}).
\btitle{Efficient pattern recognition using a new transformation distance}.
In \bbooktitle{Advances in Neural Information Processing Systems}
\bpages{50--58}.
\bpublisher{Morgan Kaufmann}, \baddress{San Mateo, CA}.
\bptok{imsref}%
\end{binproceedings}
%
\endbibitem

\bibitem[\protect\citeauthoryear{Takigawa, Kudo and Nakamura}{2009}]{Takigawa}
%
\begin{barticle}[author]
\bauthor{\bsnm{Takigawa},~\bfnm{Ichigaku}\binits{I.}},
\bauthor{\bsnm{Kudo},~\bfnm{Mineichi}\binits{M.}} \AND
\bauthor{\bsnm{Nakamura},~\bfnm{Atsuyoshi}\binits{A.}}
(\byear{2009}).
\btitle{Convex sets as prototypes for classifying patterns}.
\bjournal{Eng. Appl. Artif. Intell.}
\bvolume{22}
\bpages{101--108}.
\bptok{imsref}%
\end{barticle}
%
\endbibitem

\bibitem[\protect\citeauthoryear{Tibshirani et~al.}{2002}]{THNC2002}
%
\begin{barticle}[pbm]
\bauthor{\bsnm{Tibshirani},~\bfnm{Robert}\binits{R.}},
\bauthor{\bsnm{Hastie},~\bfnm{Trevor}\binits{T.}},
\bauthor{\bsnm{Narasimhan},~\bfnm{Balasubramanian}\binits{B.}} \AND
\bauthor{\bsnm{Chu},~\bfnm{Gilbert}\binits{G.}}
(\byear{2002}).
\btitle{Diagnosis of multiple cancer types by shrunken centroids of gene
expression}.
\bjournal{Proc. Natl. Acad. Sci. USA}
\bvolume{99}
\bpages{6567--6572}.
\bid{doi={10.1073/pnas.082099299}, issn={0027-8424}, pii={99/10/6567},
pmcid={124443}, pmid={12011421}}
\bptok{imsref}%
\end{barticle}
%
\endbibitem

\bibitem[\protect\citeauthoryear{Tipping and Sch{\"{o}}lkopf}{2001}]{Tipping01}
%
\begin{bmisc}[author]
\bauthor{\bsnm{Tipping},~\bfnm{M.~E.}\binits{M.~E.}} \AND
\bauthor{\bsnm{Sch{\"{o}}lkopf},~\bfnm{B.}\binits{B.}}
(\byear{2001}).
\bhowpublished{A kernel approach for vector quantization with guaranteed distortion
bounds. In
\textit{Artificial Intelligence and Statistics}
(T. Jaakkola and T. Richardson, eds.)
{129--134}. Morgan Kaufmann, San Francisco.}
\bptok{imsref}%
\end{bmisc}
%
\endbibitem

\bibitem[\protect\citeauthoryear{Vazirani}{2001}]{Vazirani01}
%
\begin{bbook}[mr]
\bauthor{\bsnm{Vazirani},~\bfnm{Vijay~V.}\binits{V.~V.}}
(\byear{2001}).
\btitle{Approximation Algorithms}.
\bpublisher{Springer}, \baddress{Berlin}.
\bid{mr={1851303}}
\bptok{imsref}%
\end{bbook}
%
\endbibitem

\bibitem[\protect\citeauthoryear{Wilson and Martinez}{2000}]{Wilson00}
%
\begin{barticle}[author]
\bauthor{\bsnm{Wilson},~\bfnm{D.~R.}\binits{D.~R.}} \AND
\bauthor{\bsnm{Martinez},~\bfnm{T.~R.}\binits{T.~R.}}
(\byear{2000}).
\btitle{Reduction techniques for instance-based learning algorithms}.
\bjournal{Machine Learning}
\bvolume{38}
\bpages{257--286}.
\bptok{imsref}%
\end{barticle}
%
\endbibitem

\bibitem[\protect\citeauthoryear{Zhu et~al.}{2004}]{Zhu04}
%
\begin{bincollection}[author]
\bauthor{\bsnm{Zhu},~\bfnm{Ji}\binits{J.}},
\bauthor{\bsnm{Rosset},~\bfnm{Saharon}\binits{S.}},
\bauthor{\bsnm{Hastie},~\bfnm{Trevor}\binits{T.}} \AND
\bauthor{\bsnm{Tibshirani},~\bfnm{Rob}\binits{R.}}
(\byear{2004}).
\btitle{1-norm support vector machines}.
In \bbooktitle{Advances in Neural Information Processing Systems 16}
(\beditor{\bfnm{Sebastian}\binits{S.}~\bsnm{Thrun}},
\beditor{\bfnm{Lawrence}\binits{L.}~\bsnm{Saul}} \AND
\beditor{\bfnm{Bernhard}\binits{B.}~\bsnm{{Sch{\"{o}}lkopf}}}, eds.).
\bpublisher{MIT Press}, \baddress{Cambridge, MA}.
\bptok{imsref}%
\end{bincollection}
%
\endbibitem

\end{thebibliography}
\end{document}